\documentclass[preprint,12pt]{elsarticle}
\usepackage{amssymb}
\usepackage{graphicx}
\usepackage{amssymb}
\usepackage{mathtools}
\journal{Physics Letters A}
\begin{document}
\begin{frontmatter}
\title{Shear Viscosity in the Strong Interaction Regime of a $p$-wave Superfluid Fermi Gas}
\author[A1]{Seyed Mostafa Moniri\corref{C1}}
\address[A1]{Faculty of Science, Golpayegan University of Technology, P.O. Box 87717-67498, Golpayegan, Iran}
\ead{s.m.moniri@gmail.com}
\cortext[C1]{corresponding author}
\author[A2]{Heshmatollah  Yavari}
\address[A2]{Department of Physics, University of Isfahan, Isfahan 81746, Iran}
\author[A3]{Elnaz Darsheshdar}
\address[A3]{Departamento de F\'{i}sica, Universidade Federal de S\~{a}o Carlos,
P.O. Box 676, 13565-905, S\~{a}o Carlos, S\~{a}o Paulo, Brazil}
\begin{abstract}
The $p$-wave superfluid state is a promising spin-triplet and non $s$-wave pairing state in an ultracold Fermi gas. In this work we study the low-temperature shear viscosity of a one-component $p$-wave superfluid Fermi gas, by means of Kubo formalism. 
Our study is done in the strong-coupling limit where Fermi superfluid reduces into a system of composite bosons.
Taking into account ${{p}_{x}}$-wave Cooper channel in the self-energy, the viscous relaxation rates are determined. 
The relaxation rates related to these interactions are calculated as a function of temperature. Their temperature dependence is different from the s-wave superfluid Fermi gas, and this is due to the anisotropic pairing interaction in the $p$-wave superfluid. Our results contribute to understand how this anisotropy affects transport properties of this unconventional superfluid Fermi gas in low temperature limit. 
\end{abstract}
\begin{keyword}
$P$-wave superfluid Fermi gas \sep T-matrix approximation \sep Shear viscosity
\end{keyword}
\end{frontmatter}
\section{Introduction}\label{intro}
As a spin-triplet pairing superfluid, $p$-wave superfluid Fermi gas attracts more attention during last years and was explored in both experimental \cite{Regal2003let,Regal2003nat,Ticknor2004,Zhang2004,Schunck2005,Gunter2005,Gaebler2007,Inada2008,Fuchs2008,Nakasuji2013,Maier2010} and theoretical side \cite{Gurarie2005,Ohashi2005,Ho2005,Botelho2005,Iskin2005,Iskin2006,Cheng2005,Levinsen2007,Gurarie2007,Grosfeld2007,Iskin2008,Mizushima2008,Han2009,Mizushima2010,Inotani2012}. In the Feshbach resonance of these systems, the strength of a $p$-wave interaction can be tuned from a weak-coupling to a strong-coupling regimes by controlling an external magnetic field. This was studied experimentally in $^{40}$K \cite{Regal2003let,Ticknor2004} and  $^{6}$Li \cite{Zhang2004,Schunck2005} Fermi gases. 
The weak-coupling limit is characterized by $(p_F^3 v_i)^{-1}\lesssim 0$ and the strong-coupling by $(p_F^3 v_i)^{-1}\gtrsim 0$, where $p_F$ is the Fermi momentum and $v_i (i=x,y,z)$ is the $p_i$-wave scattering volume  \cite{Inotani2015}. 
Considering the fact that the experimental realizations have been done on $p$-wave Feshbach molecules \cite{Regal2003nat,Zhang2004,Gaebler2007,Inada2008,Fuchs2008,Nakasuji2013}, in order to achieve molecules of $p$-wave interaction, it is needed to overcome a bunch of problems such as the dipolar relaxation \cite{Zhang2004,Gaebler2007,Chevy2005}, and the three-body loss \cite{Jona2008,Levinsen2008,Levinsen2007}. \\
On the other hand, applying an external magnetic field in x direction leads to the splitting of the $p$-wave Feshbach resonance into a $p_x$-wave channel and degenerate $p_y$-wave and $p_z$-wave channels by a magnetic dipole-dipole interaction. This uniaxial anisotropy is studied in a $^{40}$K Fermi gas \cite{Ticknor2004}.
It has been shown that such an anisotropic $p$-wave pairing interaction may lead to two superfluid phases with different $p$-wave pairing symmetries \cite{Gurarie2005,Gurarie2007}. The observed resonance field of a $p_x$-wave Feshbach resonance is higher than that of the other channels \cite{Ticknor2004}. It leads stronger $p_x$-wave pairing interaction ($U_x$) comparing to $p_y$-wave ($U_y$) and $p_z$-wave ($U_z$) ones, consequently, the highest superfluid phase transition temperature $T_C^{p_x}$ for $p_x$-wave superfluid phase. 
Specific condition for this uniaxial anisotropy is discussed in \cite{Gurarie2005,Gurarie2007} noticing that the system has another phase transition from the $p_x$-wave state to the $p_x+ip_y$-wave state at $T_C^{p_x+ip_y}(<T_C^{p_x})$. This indicates that plural superfluid phases can be studied  in realization of a $p$-wave superfluid Fermi gas, in all coupling ranges from the weak to the strong ones.
In addition to plural superfluid phases, an unconventional Fermi superfluid with a nodal superfluid order parameter, in its $p_x$-wave state, has an interesting pseudogap phenomenon near $T_C^{p_x+ip_y}$. In this state, the system undergos a phase transition at $T_C^{p_x}$, but single-particle excitations are gapless in the nodal direction ($\perp p_x$) thus $p_y$-wave and $p_z$-wave pairing fluctuations exist. They will be the strongest at $T_C^{p_x+ip_y}$, consequently, far below $T_C^{p_x}$, an anisotropic pseudogap phenomenon is anticipated in the nodal direction of the $p_x$-wave superfluid order parameter near $T_C^{p_x+ip_y}$. This fluctuation is expected only for non-zero $p_y$-wave and a $p_z$-wave interaction. \\
The above explained novel phenomena of an unconventional Fermi superfluid with a nodal superfluid order parameter in the $p_x$-wave state, are encouraging to investigate the shear viscosity in this system. Transport properties in the condensed matter systems have been attracted attentions continuously \cite{Ashurst1975,Ashurst2010,Enss2012,Schafer2012,Bruun2007,Taylor2010,Bobrov2010,Enss2011,Bradlyn2012,Yavari2011,Das2016,Darsheshdar2016ann,Darsheshdar2016eur,Kagamihara2019}, and was studied using Boltzmann equation approach in many works \cite{Ashurst1975,Ashurst2010,Schafer2012,Enss2012}, while the linear response theory and Kubo formula approach has attracted much more attentions \cite{Bruun2007,Taylor2010,Bobrov2010,Enss2011,Bradlyn2012,Yavari2011,Das2016,Darsheshdar2016ann,Darsheshdar2016eur,Kagamihara2019}. The superiority of the Kubo formula is due to its direct relation to diagrammatic Green’s function, thus a good control and using proper limits can be achieved. For example using fast calculation techniques for scattering in the T-matrix method \cite{Yan2008}, one can do the calculations, or in the low temperature limits, proper approximations can be applied \cite{Yavari2011}.
In the high temperature and weak interaction limits, the obtained results from Kubo formula will converge to the Boltzmann equation ones. In Boltzmann equation, it is assumed that quasiparticles having a well-defined energy-momentum relation. On the other hand, in the case of a broadened spectral functions and not well-defined quasiparticles, the effect of this broadening  and the related kinetic equation should be considered in the calculation of the transport properties. Due to the absence of a small parameter in the strong-coupling limit, this is too far ambitious. So, applying Kubo formula will be required. \\

To the best of our knowledge the shear viscosity in the strong interaction regime of a ${{p}_{x}}$-wave superfluid Fermi gas using the self-energy and the T-matrix approximation in the Kubo formula method has not been studied.
In this paper following the method of Ref. \cite{Yavari2011}, we study the low temperature shear viscosity of one component  ${{p}}$-wave superfluid Fermi gas with a uniaxially anisotropic ${{p}_{x}}$-wave pairing interaction $(U_x>U_y=U_z)$ in the strong-coupling limit. We use many-body methods to study the self-energy of the system in the T-matrix approximation (TMA). To calculate the self-energy of the system, the ${{p}_{x}}$-wave interactions are taken into account in ${{p}_{x}}$-wave Cooper channel in superfluid phases. 
\section{ Formulation }\label{sec:1}
The Shear viscosity using the Kubo formula is expressed by \cite{Mahan}, 
\begin{eqnarray}  \label{eq:1} 
\eta =-\underset{\omega \to 0}{\mathop{\lim}}
\left[ \frac{\operatorname{Im}{{\Xi }_{ret}}\left( \omega  \right)}{\omega } \right]
\end{eqnarray}
and the quantity of $\Xi_{ret} \left( {{\omega }_{n}} \right)$ named current-current correlation function and can be obtained from \cite{Mahan}, 
\begin{eqnarray}  \label{eq:2} 
\Xi \left( i{{\omega }_{n}} \right)=-\frac{1}{V}\int_{0}^{\beta }{d\tau \,{{e}^{i{{\omega }_{n}}\tau }}}\left\langle {{T}_{\tau }}S\left( \beta  \right){{e}^{{{H}_{0}}\tau }}{{\Pi }_{xy}}{{e}^{-{{H}_{0}}\tau }}{{\Pi }_{xy}} \right\rangle 
\end{eqnarray}
where $S\left( \beta  \right)$ is the S-matrix of the perturbed part of the Hamiltonian \cite{Mahan}.
$\Xi_{ret} \left( {{\omega }} \right)$ is obtained by letting $ i{{\omega }_{n}} \rightarrow  \omega+i{{\delta }} $ in $\Xi \left( i{{\omega }_{n}} \right)$ ($\delta$ is an infinitesimal parameter), and we have 
\begin{eqnarray}  \label{eq:4} 
\eta =\frac{\beta mn{{\nu }^{2}}}{16}\int{d{{\varepsilon }_{p}}}\frac{d\varepsilon }{2\pi }\,{{\operatorname{sech}}^{2}}\left( \frac{\beta \varepsilon }{2} \right)A{{\left( \mathbf{p} , \varepsilon  \right)}^{2}}
\end{eqnarray}
where the spectral function can be approximated as \cite{Mahan}
\begin{eqnarray}  \label{eq:6} 
A{{\left( \mathbf{p} , \varepsilon  \right)}^{2}}\approx \frac{2\pi \delta \left( \varepsilon -{{E}_{p}} \right)}{\operatorname{Im}\Sigma \left( \mathbf{p} , \varepsilon  \right)}
\end{eqnarray}
and viscous relaxation rate can be then defined as 
\begin{eqnarray}  \label{eq:7} 
\tau _{\eta }^{-1}=\left| 2\operatorname{Im}\Sigma \left( \mathbf{p} , \tau  \right) \right|
\end{eqnarray}
In the strong-coupling region a Feshbach resonance leads to tightly bound dimer molecules. Accordingly, in this paper we analyze the strong-coupling side of the crossover where fermions get rearranged into a system of composite bosons (dimers). The ${{p}_{x}}$-wave  superfluid state, is a spin-triplet pairing state, thus even in one component superfluid Fermi gas, fermions will behave as dimers in the strong-coupling limit. We consider specifically the limit of low temperature, whereby most of dimers are condensed. 
The Hamiltonian of one component superfluid Fermi gas with a uniaxially anisotropic ${{p}_{x}}$-wave interaction can be written as \cite{Ohashi2005},
\begin{eqnarray} \label{eq:8}
&&H=\sum\limits_{\mathbf{p}}{{}}{{\xi }_{\mathbf{p}}}c_{\mathbf{p}}^{\dagger }{{c}_{\mathbf{p}}}-\frac{1}{2}\sum\limits_{\mathbf{p} , \mathbf{{p}'} , \mathbf{q},i}{{}}{{F}_{n}}\left( \mathbf{p} \right){{p}_{i}}{{U}_{i}}{p}'_{i}{{F}_{n}}{{\left( {\mathbf{{p}'}} \right)}}\\ \nonumber
&&c_{\mathbf{p}+\mathbf{q}/2}^{\dagger }c_{-\mathbf{p}+\mathbf{q}/2}^{\dagger }{{c}_{-\mathbf{{p}'}+\mathbf{q}/2}}{{c}_{\mathbf{{p}'}+\mathbf{q}/2}}
\end{eqnarray}
where, $c_{\mathbf{p}}^{\dagger }$ is the creation operator of a Fermi atom and the kinetic energy ${{\xi }_{\mathbf{p}}}={{\varepsilon }_{\mathbf{p}}}-\mu ={{p}^{2}}/2m-\mu $, is conventionally measured relative to Fermi chemical potential $\mu $ where $m$ is the atomic mass. 
The pairing interaction in the ${{p}_{i}}$-wave Cooper channel $\left( i=x,y,z \right)$ is ${{U}_{i}}$ and we consider uniaxial anisotropy $\left( {{U}_{x}}>{{U}_{y}}={{U}_{z}} \right)$ in order to study ${{p}_{x}}$-wave superfluid with the highest superfluid phase transition
temperature. Here, we do not consider $p$-wave Feshbach resonance, but we treat $U_x$, $U_y$ and $U_z$ as tunable parameters \cite{Ticknor2004}.\\
A cutoff function is deffined in (\ref{eq:8}) in order to solve ultraviolet divergency as   ${{F}_{n}}(p)=\text{ }1/\left[ 1\text{ }+{{(p/{{p}_{c}})}^{2n}} \right]$ with $n=3$, where ${{p}_{c}}$ is a cutoff momentum. The $p$-wave interaction strength is conveniently measured in terms of the scattering volume ${{v}_{i}}$ and the effective range ${{k}_{0}}$, respectively
are given by, \cite{Inotani2012,Inotani2012low,Inotani2015,Inotani2018}, 
\begin{eqnarray} \label{eq:9}
\frac{4\pi {{v}_{i}}}{m}=-\frac{{{U}_{i}}}{3}\frac{1}{1-\frac{{{U}_{i}}}{3}\sum\limits_{p}{\frac{{{p}^{2}}}{2{{\varepsilon }_{p}}}F_{n}^{2}\left( p \right)}}
\end{eqnarray}
\begin{eqnarray} \label{eq:10}
{{k}_{0}}=-\frac{4\pi }{{{m}^{2}}}\sum\limits_{p}{\frac{{{p}^{2}}}{2\varepsilon _{p}^{2}}F_{n}^{2}\left( p \right)}
\end{eqnarray}
In the strong-coupling side in the superfluid phase, it is convenient to rewrite the Hamiltonian of Eq. (\ref{eq:8}) into the sum of the mean-field BCS part and the fluctuation corrections term. It is useful to introduce the Nambu representation for the field operators \cite{Schrieffer},
\begin{equation} \label{eq:11}
\psi_{\mathbf{p}}=\left(\begin{matrix}
&c_{\mathbf{p}}\\ 
&c_{\mathbf{p}}^{\dagger}
\end{matrix}\right)
\end{equation}
 In Nambu formalism \cite{Schrieffer}, we can rewrite Hamiltonian in Eq. (\ref{eq:8}) as below, 
\begin{eqnarray} \label{eq:12}
H=\frac{1}{2}\sum\limits_{\mathbf{p}}{{}}\psi _{\mathbf{p}}^{\dagger }\left( {{\xi }_{\mathbf{p}}}{{\tau }_{3}}-\hat{\Delta }\left( \mathbf{p} \right) \right){{\psi }_{\mathbf{p}}}-\frac{1}{2}\sum\limits_{\mathbf{q},i}{{}}{{U}_{i}}\rho _{i}^{+}\left( \mathbf{q} \right)\rho _{i}^{-}\left( -\mathbf{q} \right)
\end{eqnarray}
where ${{\tau }_{i}}\left( i=x,y,z \right)$ are Pauli matrices. The first term in Eq. (\ref{eq:12}) is just the mean-field BCS Hamiltonian, where 
\begin{eqnarray} \label{eq:13}
\hat{\Delta }\left( \mathbf{p} \right)=\left( \begin{matrix} 
   0 & \Delta \left( \mathbf{p} \right)  \\ 
   {{\Delta }^{*}}\left( \mathbf{p} \right) & {}  \\ \end{matrix} \right)
\end{eqnarray}
is a $p$-wave superfluid order parameter and,
\begin{eqnarray} \label{eq:14}
\Delta \left( \mathbf{p} \right)=\mathbf{b}.\mathbf{p}{{F}_{n}}\left( \mathbf{p} \right)
\end{eqnarray}
where $\mathbf{b}=\left( {{b}_{x}},{{b}_{y}},{{b}_{z}} \right)$ has the form of
\begin{eqnarray} \label{eq:15}
{{b}_{i}}={{U}_{i}}\sum\limits_{{}}{{{p}_{i}}{{F}_{n}}\left( \mathbf{p} \right)\left\langle {{c}_{-\mathbf{p}}}{{c}_{\mathbf{p}}} \right\rangle }
\end{eqnarray}
The $\rho _{i}^{\pm }\left( \mathbf{q} \right)$ is a generalized density operator for a pair of fermions that physically describes ${{p}_{i}}$-wave superfluid fluctuations and can be written as 
\begin{eqnarray} \label{eq:16}
\rho _{i}^{\pm }\left( \mathbf{q} \right)=\sum\limits_{{}}{{{p}_{i}}{{F}_{n}}\left( \mathbf{p} \right)\psi _{\mathbf{p}+\mathbf{q}/2}^{\dagger }{{\tau }_{\pm }}{{\psi }_{\mathbf{p}-\mathbf{q}/2}}}
\end{eqnarray}
and ${{\tau }_{\pm }}\text{ }=\text{ }\left( {{\tau }_{x}}\pm {{\tau}_{y}} \right)/2$. The second term in Eq. (\ref{eq:12}) gives fluctuation corrections to the mean-field Hamiltonian. The BCS mean-field Green’s function is,
\begin{eqnarray} \label{eq:17}
{{\hat{G}}_{0}}\left( \mathbf{p},i{{\omega }_{m}} \right)=\frac{1}{i{{\omega }_{m}}-{{\xi }_{\mathbf{p}}}{{\tau }_{3}}+\hat{\Delta }\left( \mathbf{p} \right)}
\end{eqnarray}
where ${{\omega }_{m}}$ is the fermionic Matsubara frequency. Fluctuation corrections to the single-particle mean-field Hamiltonian can be conveniently considered by the self-energy $\Sigma \left( \mathbf{p},i{{\omega }_{m}} \right)$ in the single-particle thermal Green’s function,
\begin{eqnarray} \label{eq:18}
\hat{G}\left( \mathbf{p},i{{\omega }_{m}} \right)=\frac{1}{\hat{G}_{0}^{-1}\left( \mathbf{p},i{{\omega }_{m}} \right)-\hat{\Sigma }\left( \mathbf{p},i{{\omega }_{m}} \right)}
\end{eqnarray}
\begin{figure}
\centering
\includegraphics[width=0.8\textwidth]{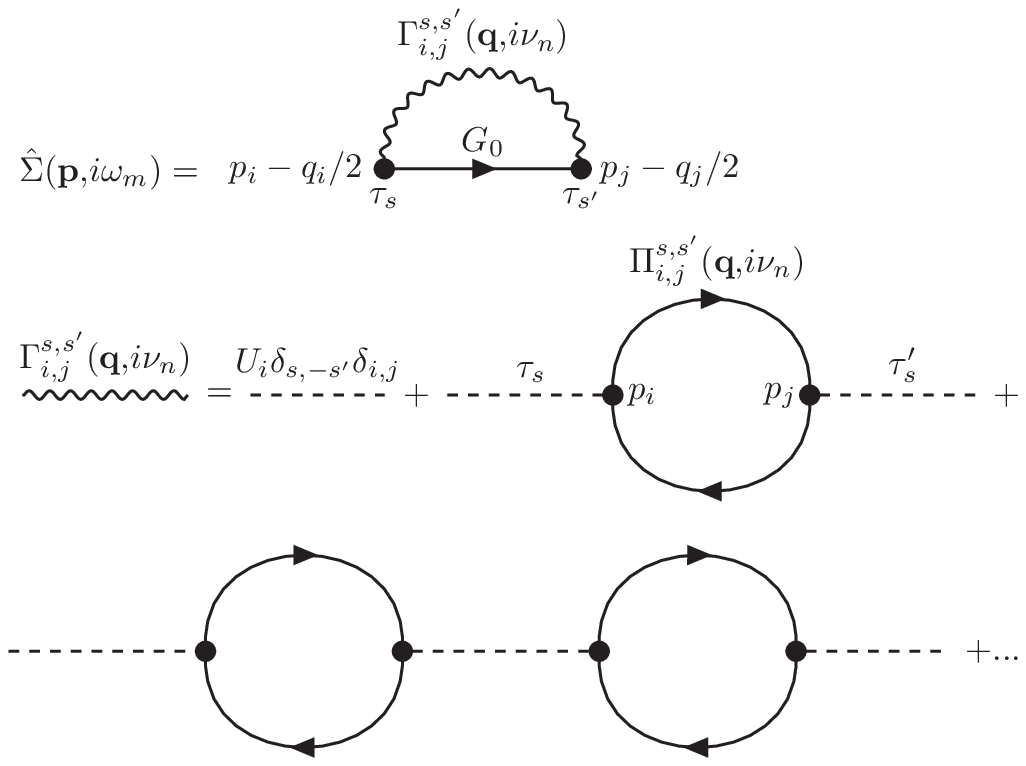}
\caption{Feynman diagrams describing the self-energy $\hat{\Sigma} \left( \mathbf{p},i{{\omega }_{m}} \right)$ and particle-particle scattering matrix $\Gamma _{i,j}^{s,{s}'}\left( \mathbf{q},i{{\nu }_{n}} \right)$ that is shown by wavy line in the TMA. The solid line and the dashed line describe the mean-field single-particle Green’s function $\hat{G}_{0}\left( \mathbf{p},i{{\omega }_{m}} \right)$ and the p-wave interaction, respectively. }
\label{fig:1}
\end{figure}
The self-energy $\Sigma \left( \mathbf{p},i{{\omega }_{m}} \right)$ within the TMA can be written as, 
\begin{eqnarray} \label{eq:19}
&& \hat{\Sigma }\left( \mathbf{p},i{{\omega }_{n}} \right)=\frac{2}{\beta }\sum\limits_{i,j}{{}}\sum\limits_{\mathbf{q},i{{\nu }_{n}}}{{}}F_{n}^{2}\left( \mathbf{p}-\frac{\mathbf{q}}{2} \right)\left[ {{p}_{i}}-\frac{{{q}_{i}}}{2} \right]\left[ {{p}_{j}}-\frac{{{q}_{j}}}{2} \right] \\ \nonumber
 && \left( \begin{matrix}  
   G_{0}^{22}\left( \mathbf{p}-\mathbf{q},i{{\omega }_{m}}-i{{\nu }_{n}} \right)\Gamma _{i,j}^{-+}\left( \mathbf{q},i{{\nu }_{n}}  \right) & G_{0}^{21}\left( \mathbf{p}-\mathbf{q},i{{\omega }_{m}}-i{{\nu }_{n}} \right)\Gamma _{i,j}^{--}\left( \mathbf{q},i{{\nu }_{n}} \right)  \\  
   G_{0}^{12}\left( \mathbf{p}-\mathbf{q},i{{\omega }_{m}}-i{{\nu }_{n}} \right)\Gamma _{i,j}^{++}\left( \mathbf{q},i{{\nu }_{n}} \right) & G_{0}^{11}\left( \mathbf{p}-\mathbf{q},i{{\omega }_{m}}-i{{\nu }_{n}} \right)\Gamma _{i,j}^{+-}\left( \mathbf{q},i{{\nu }_{n}} \right) 
\end{matrix} \right) 
\end{eqnarray}
The self-energy $\Sigma \left( \mathbf{p},i{{\omega }_{m}} \right)$ is diagrammatically given in Fig. (\ref{fig:1}). The TMA particle-particle scattering matrix $\Gamma _{i,j}^{s,{s}'}\left( \mathbf{q},i{{\nu }_{n}} \right)$ gives the equation,
\begin{eqnarray} \label{eq:20}
\Gamma _{i,j}^{s,{s}'}\left( \mathbf{q},i{{\nu }_{n}} \right)=-{{U}_{i}}{{\delta }_{i,j}}{{\delta }_{s,-{s}'}}-{{U}_{i}}\sum\limits_{{{s}'}'=\pm }{{}}\sum\limits_{k}{{}}\Pi _{i,k}^{s,{{s}'}'}\left( \mathbf{q},i{{\nu }_{n}} \right)\Gamma _{k,j}^{-{{s}'}',{s}'}\left( \mathbf{q},i{{\nu }_{n}} \right)
\end{eqnarray}
and,
\begin{eqnarray} \label{eq:21}
&&\Pi _{i,j}^{s,{s}'}\left( \mathbf{q},i{{\nu }_{n}} \right)=\frac{1}{\beta }\sum\limits_{\mathbf{p},i{{\omega }_{m}}}{{}}{{p}_{i}}{{p}_{j}}F_{n}^{2}\left( p \right)\text{Tr}\\ \nonumber
&&\left[ {{\tau }_{s}}{{G}_{0}}\left( \mathbf{p}+\mathbf{q}/2,i{{\omega }_{n}} \right){{\tau }_{{{s}'}}}{{G}_{0}}\left( \mathbf{p}-\mathbf{q}/2,i{{\omega }_{n}}-i{{\nu }_{n}} \right) \right]
\end{eqnarray}
is the pair correlation function. In particular, $\Pi _{i,i}^{s,{s}'}\left( \mathbf{q},i{{\nu }_{n}} \right)$ describes fluctuations in the ${{p}_{i}}$-wave Cooper channel, and $\Pi _{i,j}^{s,{s}'}\left( \mathbf{q},i{{\nu }_{n}} \right)\left( i\ne j \right)$ describes coupling between ${{p}_{i}}$-wave and ${{p}_{j}}$-wave pairing fluctuations. We performed the Matsubara summation, for the self-energy $\hat{\Sigma }\left( \mathbf{p},i{{\omega }_{n}} \right)$ in Eq. (\ref{eq:19}), it gives, 
\begin{eqnarray} \label{eq:22}
 && {{\Sigma }_{11}}\left( \mathbf{p},i{{\omega }_{m}} \right)=2\sum\limits_{i,j}{{}}\sum\limits_{\mathbf{q}}{{}}F_{n}^{2}\left( \mathbf{p}-\frac{\mathbf{q}}{2} \right)\left[ {{p}_{i}}-\frac{{{q}_{i}}}{2} \right]\left[ {{p}_{j}}-\frac{{{q}_{j}}}{2} \right] \\ \nonumber
 && \bigg( \left[ \frac{1}{2}-\frac{{{\xi }_{\mathbf{p}-\mathbf{q}}}}{2{{E}_{\mathbf{p}-\mathbf{q}}}}\tanh \left( \frac{\beta }{2}{{E}_{\mathbf{p}-\mathbf{q}}} \right) \right]- \\ \nonumber
 &&\int{{}}\frac{d\varepsilon }{2\pi }{{n}_{B}}\left( \varepsilon  \right)B_{i,j}^{-+}\left( \mathbf{p}-\mathbf{q} , \varepsilon  \right)G_{22}^{0}\left( \mathbf{p}-\mathbf{q},i{{\omega }_{m}}-\varepsilon  \right) \bigg)
\end{eqnarray}
and,
\begin{eqnarray} \label{eq:23}
&& {{\Sigma }_{21}}\left( \mathbf{p},i{{\omega }_{m}} \right)=2\sum\limits_{i,j}{{}}\sum\limits_{\mathbf{q}}{{}}F_{n}^{2}\left( \mathbf{p}-\frac{\mathbf{q}}{2} \right)\left[ {{p}_{i}}-\frac{{{q}_{i}}}{2} \right]\left[ {{p}_{j}}-\frac{{{q}_{j}}}{2} \right] \\ \nonumber
&& \Delta \left( \mathbf{p}-\mathbf{q} \right) \left( \frac{1}{2{{E}_{\mathbf{p}-\mathbf{q}}}}-\int{{}}\frac{d\varepsilon }{2\pi }{{n}_{B}}\left( \varepsilon  \right)B_{i,j}^{++}\left( \mathbf{q} , \varepsilon  \right)G_{12}^{0}\left( \mathbf{p}-\mathbf{q},i{{\omega }_{m}}-\varepsilon  \right) \right)
\end{eqnarray}
where ${{n}_{B}}\left( \varepsilon  \right)$ and ${{n}_{F}}\left( \varepsilon  \right)$ are the bosonic and fermionic distribution functions respectively, ${{E}_{\mathbf{p}}}=\sqrt{\xi _{\mathbf{p}}^{2}+{{\Delta }^{2}}\left( \mathbf{p} \right)}$ is the Bogoliubov single-particle excitation spectrum and $B_{i,j}^{s,{s}'}\left( \mathbf{q} , \varepsilon  \right)=-2\operatorname{Im}\Gamma _{i,j}^{s,{s}',\text{ret}}\left( \mathbf{q} , \varepsilon  \right)$. The order parameter $\Delta \left( \mathbf{p} \right)$ is related to the ${{p}_{x}}$-wave superfluid phase \cite{Inotani2015} and it differs from the ${s}$-wave one by its momentum dependency. The other components of self-energy can be obtained by using $\Pi _{i,j}^{--}\left( \mathbf{q},i{{\nu }_{n}} \right)=\Pi _{i,j}^{++}\left( \mathbf{q},i{{\nu }_{n}} \right)$ and $\Pi _{i,j}^{+-}\left( \mathbf{q},i{{\nu }_{n}} \right)=\Pi _{i,j}^{-+}\left( \mathbf{q},-i{{\nu }_{n}} \right)$.
In the TMA for the particle-particle scattering matrix $\Gamma _{i,j}^{s,{s}'}\left( \mathbf{q},i{{\nu }_{n}} \right)$ we consider the first two terms as,
\begin{eqnarray} \label{eq:24}
\Gamma _{i,j}^{s,{s}'}\left( \mathbf{q},i{{\nu }_{n}} \right)=-{{U}_{i}}{{\delta }_{i,j}}{{\delta }_{s,-{s}'}}+{{U}_{i}}{{U}_{j}}\Pi _{i,j}^{s,{s}'}\left( \mathbf{q},i{{\nu }_{n}} \right)
\end{eqnarray}
The momentum dependent nature of the interaction in $p$-wave superfluids, allows us to use this approximation in low momentum limits, in order to remove higher power of the momentum.  
The pair correlation function, $\Pi _{i,j}^{s,{s}'}\left( \mathbf{q},i{{\nu }_{n}} \right)$, can be obtained after doing the Matsubara summation in Eq. (\ref{eq:21}) as below, 
\begin{eqnarray} \label{eq:25}
&& \Pi _{i,j}^{++}\left( \mathbf{q},i{{\nu }_{n}} \right)=  \frac{1}{4}\sum\limits_{\mathbf{p}}{{}}{{p}_{i}}{{p}_{j}}F_{n}^{2}\left( p \right)\frac{\Delta _{\mathbf{p}+\mathbf{q}/2}^{*}\Delta _{\mathbf{p}-\mathbf{q}/2}^{*}}{{{E}_{\mathbf{p}+\mathbf{q}/2}}{{E}_{\mathbf{p}-\mathbf{q}/2}}} \\  \nonumber
&&\Bigg( \frac{{{E}_{\mathbf{p}+\mathbf{q}/2}}+{{E}_{\mathbf{p}-\mathbf{q}/2}}}{\left( {{\nu }^{2}}_{n}+{{\left( {{E}_{\mathbf{p}+\mathbf{q}/2}}+{{E}_{\mathbf{p}-\mathbf{q}/2}} \right)}^{2}} \right)}\left[ \tanh \left( \frac{\beta }{2}{{E}_{\mathbf{p}+\mathbf{q}/2}} \right)+\tanh \left( \frac{\beta }{2}{{E}_{\mathbf{p}-\mathbf{q}/2}} \right) \right] \\  \nonumber
 && -\frac{{{E}_{\mathbf{p}+\mathbf{q}/2}}-{{E}_{\mathbf{p}-\mathbf{q}/2}}}{\left( {{\nu }^{2}}_{n}+{{\left( {{E}_{\mathbf{p}+\mathbf{q}/2}}-{{E}_{\mathbf{p}-\mathbf{q}/2}} \right)}^{2}} \right)}\left[ \tanh \left( \frac{\beta }{2}{{E}_{\mathbf{p}+\mathbf{q}/2}} \right)-\tanh \left( \frac{\beta }{2}{{E}_{\mathbf{p}-\mathbf{q}/2}} \right) \right] \Bigg)
\end{eqnarray}
and,
\begin{eqnarray} \label{eq:26}
&& \Pi _{i,j}^{-+}\left( \mathbf{q},i{{\nu }_{n}} \right)=\frac{1}{4}\sum\limits_{\mathbf{p}}{{}}{{p}_{i}}{{p}_{j}}F_{n}^{2}\left( p \right) \\  \nonumber
 &&\Bigg( \left[ \tanh \left( \frac{\beta }{2}{{E}_{\mathbf{p}+\mathbf{q}/2}} \right)+\tanh \left( \frac{\beta }{2}{{E}_{\mathbf{p}-\mathbf{q}/2}} \right) \right] \\ \nonumber
 && \Bigg[ \left( 1+\frac{{{\xi }_{\mathbf{p}+\mathbf{q}/2}}{{\xi }_{\mathbf{p}-\mathbf{q}/2}}}{{{E}_{\mathbf{p}+\mathbf{q}/2}}{{E}_{\mathbf{p}-\mathbf{q}/2}}} \right)\frac{1}{i{{\nu }_{n}}-\left( {{E}_{\mathbf{p}+\mathbf{q}/2}}+{{E}_{\mathbf{p}-\mathbf{q}/2}} \right)} \\  \nonumber
 && +\left( 1-\frac{{{\xi }_{\mathbf{p}+\mathbf{q}/2}}}{{{E}_{\mathbf{p}+\mathbf{q}/2}}} \right)\left( 1-\frac{{{\xi }_{\mathbf{p}-\mathbf{q}/2}}}{{{E}_{\mathbf{p}-\mathbf{q}/2}}} \right)\frac{i{{\nu }_{n}}}{\nu _{n}^{2}+{{\left( {{E}_{\mathbf{p}+\mathbf{q}/2}}+{{E}_{\mathbf{p}-\mathbf{q}/2}} \right)}^{2}}} \Bigg] \\ \nonumber 
 && + \left[ \tanh \left( \frac{\beta }{2}{{E}_{\mathbf{p}+\mathbf{q}/2}} \right)-\tanh \left( \frac{\beta }{2}{{E}_{\mathbf{p}-\mathbf{q}/2}} \right) \right] \\ \nonumber
 && \Bigg[ 	\left( 1-\frac{{{\xi }_{\mathbf{p}+\mathbf{q}/2}}{{\xi }_{\mathbf{p}-\mathbf{q}/2}}}{{{E}_{\mathbf{p}+\mathbf{q}/2}}{{E}_{\mathbf{p}-\mathbf{q}/2}}} \right)\frac{1}{i{{\nu }_{n}}-\left( {{E}_{\mathbf{p}+\mathbf{q}/2}}-{{E}_{\mathbf{p}-\mathbf{q}/2}} \right)} \\  \nonumber
&& +\left( 1-\frac{{{\xi }_{\mathbf{p}+\mathbf{q}/2}}}{{{E}_{\mathbf{p}+\mathbf{q}/2}}} \right)\left( 1+\frac{{{\xi }_{\mathbf{p}-\mathbf{q}/2}}}{{{E}_{\mathbf{p}-\mathbf{q}/2}}} \right)\frac{i{{\nu }_{n}}}{\nu _{n}^{2}+{{\left( {{E}_{\mathbf{p}+\mathbf{q}/2}}-{{E}_{\mathbf{p}-\mathbf{q}/2}} \right)}^{2}}} \Bigg] \Bigg)
\end{eqnarray}
By substituting Eqs.(\ref{eq:25}) and (\ref{eq:26}) in Eq. (\ref{eq:24}) and using analytic continuation we can obtain spectral function of $B_{i,j}^{s,{s}'}\left( \mathbf{q},\varepsilon  \right)=-2\operatorname{Im}\Gamma _{i,j}^{s,{s}',\text{ret}}\left( \mathbf{q},\varepsilon  \right)$ as below, 
\begin{eqnarray} \label{eq:27}
&& B_{i,j}^{++}\left( \mathbf{q},\varepsilon  \right)=\frac{\pi }{4}{{U}_{i}}{{U}_{j}}\sum\limits_{\mathbf{p}}{{}}{{p}_{i}}{{p}_{j}}F_{n}^{2}\left( p \right)\frac{\Delta _{\mathbf{p}+\mathbf{q}/2}^{*}\Delta _{\mathbf{p}-\mathbf{q}/2}^{*}}{{{E}_{\mathbf{p}+\mathbf{q}/2}}{{E}_{\mathbf{p}-\mathbf{q}/2}}} \\  \nonumber
&&\Bigg( \left[ \delta \left( \varepsilon -{{E}_{\mathbf{p}+\mathbf{q}/2}}-{{E}_{\mathbf{p}-\mathbf{q}/2}} \right)-\delta \left( \varepsilon +{{E}_{\mathbf{p}+\mathbf{q}/2}}+{{E}_{\mathbf{p}-\mathbf{q}/2}} \right) \right]\\ \nonumber
 &&\left[ \tanh \left( \frac{\beta }{2}{{E}_{\mathbf{p}+\mathbf{q}/2}} \right)+\tanh \left( \frac{\beta }{2}{{E}_{\mathbf{p}-\mathbf{q}/2}} \right) \right] \\  \nonumber
&& +\left[ \delta \left( \varepsilon +{{E}_{\mathbf{p}+\mathbf{q}/2}}-{{E}_{\mathbf{p}-\mathbf{q}/2}} \right)-\delta \left( \varepsilon -{{E}_{\mathbf{p}+\mathbf{q}/2}}+{{E}_{\mathbf{p}-\mathbf{q}/2}} \right) \right]\\ \nonumber
 &&\left[ \tanh \left( \frac{\beta }{2}{{E}_{\mathbf{p}+\mathbf{q}/2}} \right)-\tanh \left( \frac{\beta }{2}{{E}_{\mathbf{p}-\mathbf{q}/2}} \right) \right] \Bigg)
\end{eqnarray}
and,
\begin{eqnarray} \label{eq:28}
 && B_{i,j}^{-+}\left( \mathbf{q},\varepsilon  \right)=\frac{\pi }{2}{{U}_{i}}{{U}_{j}}\sum\limits_{\mathbf{p}}{{}}{{p}_{i}}{{p}_{j}}F_{n}^{2}\left( p \right) \\ \nonumber
 && \Bigg(\Bigg[ \frac{1}{2}\left( 1-\frac{{{\xi }_{\mathbf{p}+\mathbf{q}/2}}}{{{E}_{\mathbf{p}+\mathbf{q}/2}}} \right)\left( 1-\frac{{{\xi }_{\mathbf{p}-\mathbf{q}/2}}}{{{E}_{\mathbf{p}-\mathbf{q}/2}}} \right)\delta \left( \varepsilon +{{E}_{\mathbf{p}+\mathbf{q}/2}}+{{E}_{\mathbf{p}-\mathbf{q}/2}} \right) \\  \nonumber
 && +\left( 1+\frac{{{\xi }_{\mathbf{p}+\mathbf{q}/2}}{{\xi }_{\mathbf{p}-\mathbf{q}/2}}}{{{E}_{\mathbf{p}+\mathbf{q}/2}}{{E}_{\mathbf{p}-\mathbf{q}/2}}}+\frac{1}{2}\left( 1-\frac{{{\xi }_{\mathbf{p}+\mathbf{q}/2}}} {{{E}_{\mathbf{p}+\mathbf{q}/2}}} \right)\left( 1-\frac{{{\xi }_{\mathbf{p}-\mathbf{q}/2}}}{{{E}_{\mathbf{p}-\mathbf{q}/2}}} \right) \right)\\ \nonumber
 &&\delta \left( \varepsilon -{{E}_{\mathbf{p}+\mathbf{q}/2}}-{{E}_{\mathbf{p}-\mathbf{q}/2}} \right)  \Bigg]\left[ \tanh \left( \frac{\beta }{2}{{E}_{\mathbf{p}+\mathbf{q}/2}} \right)+\tanh \left( \frac{\beta }{2}{{E}_{\mathbf{p}-\mathbf{q}/2}} \right) \right] \\ \nonumber  
 && +\Bigg[ \frac{1}{2}\left( 1-\frac{{{\xi }_{\mathbf{p}+\mathbf{q}/2}}}{{{E}_{\mathbf{p}+\mathbf{q}/2}}} \right)\left( 1+\frac{{{\xi }_{\mathbf{p}-\mathbf{q}/2}}}{{{E}_{\mathbf{p}-\mathbf{q}/2}}} \right)\delta \left( \varepsilon +{{E}_{\mathbf{p}+\mathbf{q}/2}}-{{E}_{\mathbf{p}-\mathbf{q}/2}} \right) \\  \nonumber
 && +\left( 1-\frac{{{\xi }_{\mathbf{p}+\mathbf{q}/2}}{{\xi }_{\mathbf{p}-\mathbf{q}/2}}}{{{E}_{\mathbf{p}+\mathbf{q}/2}}{{E}_{\mathbf{p}-\mathbf{q}/2}}}+\frac{1}{2}\left( 1-\frac{{{\xi }_{\mathbf{p}+\mathbf{q}/2}}}{{{E}_{\mathbf{p}+\mathbf{q}/2}}} \right)\left( 1+\frac{{{\xi }_{\mathbf{p}-\mathbf{q}/2}}}{{{E}_{\mathbf{p}-\mathbf{q}/2}}} \right) \right)\\ \nonumber
 &&\delta \left( \varepsilon -{{E}_{\mathbf{p}+\mathbf{q}/2}}+{{E}_{\mathbf{p}-\mathbf{q}/2}} \right) \Bigg]\left[ \tanh \left( \frac{\beta }{2}{{E}_{\mathbf{p}+\mathbf{q}/2}} \right)-\tanh \left( \frac{\beta }{2}{{E}_{\mathbf{p}-\mathbf{q}/2}} \right) \right] \Bigg)
\end{eqnarray}
As we have the anisotropy of $U_x > U_y = U_z$, the superfluid phase is dominated by the $p_x$-wave pairing interaction in BEC limit. Then we limit our calculation to $p_x$-wave cooper channel.
We approximate $B_{xx}^{s,{s}'}\left( \mathbf{q},\varepsilon  \right)$ at low momentum and temperature. Then after integrating, we have, 
\begin{eqnarray} \label{eq:29}
&&B_{xx}^{++}\left( \mathbf{q},\varepsilon  \right)=\frac{U_{x}^{2}}{24\pi }\sqrt{{{\Delta }^{2}}\left( {{\mathbf{p}}_{0}}-\mathbf{q} \right)+{{\mu }^{2}}}{{p}_{0}}^{3}F_{n}^{2}\left( {{p}_{0}} \right)\frac{{{\Delta }^{2}}\left( {{\mathbf{p}}_{0}}-\mathbf{q} \right)}{{{E}_{{{\mathbf{p}}_{0}}+\mathbf{q}/2}}{{E}_{{{\mathbf{p}}_{0}}-\mathbf{q}/2}}}\\ \nonumber
 &&\left( 1-{{\text{e}}^{-2\beta {{E}_{{{\mathbf{p}}_{_{0}}}+\mathbf{q}/2}}}}-{{\text{e}}^{-2\beta {{E}_{{{\mathbf{p}}_{_{0}}}-\mathbf{q}/2}}}} \right)
\end{eqnarray}
and,
\begin{eqnarray} \label{eq:30}
&& B_{xx}^{-+}\left( \mathbf{q},\varepsilon  \right)=\frac{1}{12\pi }U_{x}^{2}\,\sqrt{{{\Delta }^{2}}\left( {{\mathbf{p}}_{0}}-\mathbf{q} \right)+{{\mu }^{2}}}{{p}_{0}}^{3}F_{n}^{2}\left( {{p}_{0}} \right)\\ \nonumber
 &&\left( 1-{{\text{e}}^{-2\beta {{E}_{{{\mathbf{p}}_{_{0}}}+\mathbf{q}/2}}}}-{{\text{e}}^{-2\beta {{E}_{{{\mathbf{p}}_{_{0}}}-\mathbf{q}/2}}}} \right) \\ \nonumber
&& \left( 1+\frac{{{\xi }_{{{\mathbf{p}}_{0}}+\mathbf{q}/2}}{{\xi }_{{{\mathbf{p}}_{0}}-\mathbf{q}/2}}}{{{E}_{{{\mathbf{p}}_{0}}+\mathbf{q}/2}}{{E}_{{{\mathbf{p}}_{0}}-\mathbf{q}/2}}}+\frac{1}{2}\left( 1-\frac{{{\xi }_{{{\mathbf{p}}_{0}}+\mathbf{q}/2}}}{{{E}_{{{\mathbf{p}}_{0}}+\mathbf{q}/2}}} \right)\left( 1-\frac{{{\xi }_{{{\mathbf{p}}_{0}}-\mathbf{q}/2}}}{{{E}_{{{\mathbf{p}}_{0}}-\mathbf{q}/2}}} \right) \right)
\end{eqnarray}
where ${{p}_{0}}=\sqrt{\left(2{{\Delta }^{2}}(\textbf{q})+2{{\mu }^{2}}-\varepsilon \sqrt{{{\Delta }^{2}}(\textbf{q})+{{\mu }^{2}}}\right)/\mu}$. Substituting above equations to Eq. (\ref{eq:22}), (\ref{eq:23}) and performing analytic continuation, the imaginary parts of the diagonal and the off-diagonal parts of the self-energy matrix are,
\begin{eqnarray} \label{eq:31}
&& \operatorname{Im}\left[ {{\Sigma }_{11}}\left( \mathbf{p},{{\omega }_{m}} \right) \right]=\frac{1}{2}\sum\limits_{\mathbf{q}}{{}}F_{n}^{2}\left( \mathbf{p}-\frac{\mathbf{q}}{2} \right)\left[ {{p}_{x}}-\frac{{{q}_{x}}}{2} \right]\left[ {{p}_{x}}-\frac{{{q}_{x}}}{2} \right] \\  \nonumber
&& \Bigg( \left( 1-\frac{{{\xi }_{\mathbf{p}-\mathbf{q}}}}{{{E}_{\mathbf{p}-\mathbf{q}}}} \right){{n}_{B}}\left( {{\omega }_{m}}-{{E}_{\mathbf{p}-\mathbf{q}}} \right)B_{xx}^{-+}\left( \mathbf{p}-\mathbf{q},{{\omega }_{m}}-{{E}_{\mathbf{p}-\mathbf{q}}} \right)\\  \nonumber
&&+\left( 1+\frac{{{\xi }_{\mathbf{p}-\mathbf{q}}}}{{{E}_{\mathbf{p}-\mathbf{q}}}} \right){{n}_{B}}\left( {{\omega }_{m}}+{{E}_{\mathbf{p}-\mathbf{q}}} \right)B_{xx}^{-+}\left( \mathbf{p}-\mathbf{q},{{\omega }_{m}}+{{E}_{\mathbf{p}-\mathbf{q}}} \right) \Bigg)
\end{eqnarray}
and,
\begin{eqnarray} \label{eq:32}
&& \operatorname{Im}\left[ {{\Sigma }_{21}}\left( \mathbf{p},{{\omega }_{m}} \right) \right]=\sum\limits_{\mathbf{q}}{{}}F_{n}^{2}\left( \mathbf{p}-\frac{\mathbf{q}}{2} \right)\left[ {{p}_{x}}-\frac{{{q}_{x}}}{2} \right]\left[ {{p}_{x}}-\frac{{{q}_{x}}}{2} \right] \\  \nonumber
&& \frac{{{\Delta }^{2}}\left( \mathbf{p}-\mathbf{q} \right)}{2{{E}_{\mathbf{p}-\mathbf{q}}}}\big( {{n}_{B}}\left( {{\omega }_{m}}+{{E}_{\mathbf{p}-\mathbf{q}}} \right)B_{xx}^{++}\left( \mathbf{q},{{\omega }_{m}}+{{E}_{\mathbf{p}-\mathbf{q}}} \right) \\ \nonumber
&&-{{n}_{B}}\left( {{\omega }_{m}}-{{E}_{\mathbf{p}-\mathbf{q}}} \right)B_{xx}^{++}\left( \mathbf{q},{{\omega }_{m}}-{{E}_{\mathbf{p}-\mathbf{q}}} \right) \big) 
\end{eqnarray}
Doing the integration over ${{d}^{3}}q$ in Eq. (\ref{eq:31}) and (\ref{eq:32}) in low temperature limit, one can obtain temperature behaviour of the imaginary part of the self-energy as,
\begin{eqnarray} \label{eq:33}
\operatorname{Im}\left[ {{\Sigma }_{11}}\left( \mathbf{p},{{\omega }_{m}} \right) \right] \propto e^{-E_p/k_BT}\left( T^4+T^3 \right)
\end{eqnarray}
and,
\begin{eqnarray} \label{eq:34}
\operatorname{Im}\left[ {{\Sigma }_{21}}\left( \mathbf{p},{{\omega }_{m}} \right) \right]\propto T^3 \left( e^{-E_p/k_BT} - e^{-3E_p/k_BT}\right)
\end{eqnarray}
Thus from Eqs. (\ref{eq:7}), (\ref{eq:33}) and (\ref{eq:34}), the variation of the viscous relaxation rate with temperature can be described as 
\begin{eqnarray} \label{eq:37}
\tau _{11}^{-1}\propto e^{-E_p/k_BT}\left( T^4+T^3 \right)
\end{eqnarray}
\begin{eqnarray} \label{eq:38}
\tau _{21}^{-1}\propto T^3 \left( e^{-E_p/k_BT} - e^{-3E_p/k_BT}\right)
\end{eqnarray}
By using Eqs. (\ref{eq:6}) and (\ref{eq:7}) into Eq. (\ref{eq:4}) one can write, 
\begin{eqnarray} \label{eq:39}
\eta =\frac{\beta nm{{v}^{2}}{{\tau }_{\eta }}}{8}\int{d{{\varepsilon }_{p}}}\operatorname{sech}{{\left( \frac{\beta {{E}_{p}}}{2} \right)}^{2}}
\end{eqnarray}
Integrating of Eq. (\ref{eq:39}) in the long wavelength limit $(p\to 0)$, we can write the temperature dependence of shear viscosity as, 
\begin{eqnarray} \label{eq:40}
{{\eta }_{11}}\propto e^{E_p/k_BT}\left( T^4+T^3 \right)^{-1}
\end{eqnarray}
\begin{eqnarray} \label{eq:41}
{{\eta }_{21}} \propto T^{-3} \left( e^{-E_p/k_BT} - e^{-3E_p/k_BT}\right)^{-1}
\end{eqnarray}
Low temperature behaviour of the relaxation rates is shown in Fig. (\ref{fig:2}) for ${{(p_{F}^{3}{{v}_{x}})}^{-3}}\approx 0.0$ in the strong coupling limit. In this plot we used low temperature values of the order parameter obtained in \cite{Inotani2015}. The difference in $\tau _{11}^{-1}$ and $\tau _{21}^{-1}$ plots is due to the different integrals in Eqs. (\ref{eq:22}) and (\ref{eq:23}) leading to differences in (\ref{eq:37}) and (\ref{eq:38}). The obtained temperature dependences of the relaxation rates and the shear viscosities have not been reported in the non-interacting $s$-wave superfluid Fermi gas where we used the same method \cite{Yavari2011}, this novelty is due to the uniaxially anisotropic ${{p}_{x}}$-wave pairing interaction $(U_x>U_y=U_z)$ and the momentum dependent order parameter. Although equation form of the temperature dependence is new, the general trend of the temperature behavior in the shear viscosities is similar to other superfluid Fermi gases like in Fig. (1) of ref. \cite{Schafer2012}, and in a superfluid Fermi gas in the unitarity limit \cite{Rupak2007}, as they all diverge at infinite $\beta $. 
\begin{figure}
\centering
\includegraphics[width=0.8\textwidth]{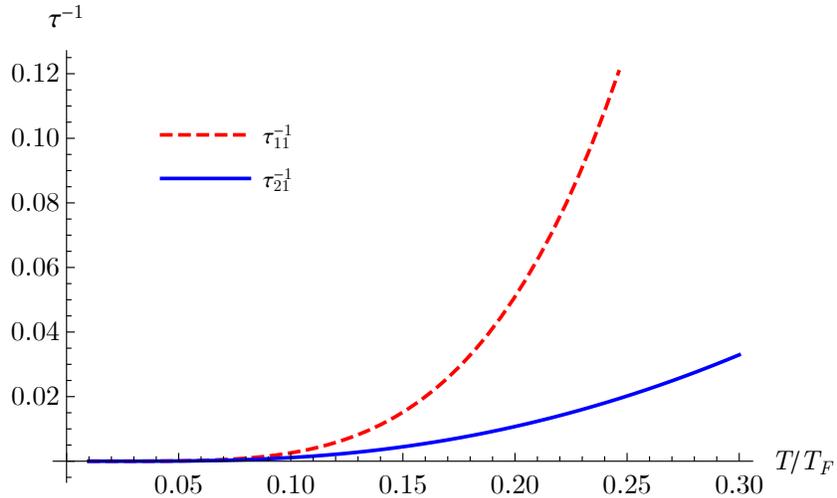}
\label{fig:2}
\caption{ The viscous relaxation rates in the unit of $\frac{{{\varepsilon }_{F}}}{\hbar }$ (${{\varepsilon }_{F}}$ is the Fermi energy and $T_F$ is the Fermi temperature).}
\end{figure}
\section{Conclusion} \label{sec:5}
Using the linear response theory and Green’s function method, shear viscosity of one component $p$-wave superfluid Fermi gas with a uniaxially anisotropic pairing interaction has been studied. The system was studied in the strong-coupling limit where fermions get rearranged into a system of composite bosons, and in the limit of low temperature, where most of composite bosons are condensed. The viscous relaxation rates (${{\tau }^{-1}}=\left| 2\operatorname{Im}\Sigma  \right|$) related to collisions in the diagonal and off-diagonal parts of the self-energy, have been calculated, taking into account the ${{p}_{x}}$-wave Cooper channel interaction.  It has been shown that the $\tau _{11}^{-1}$ viscous relaxation rate has $e^{-E_p/k_BT}\left( T^4+T^3 \right)$ temperature behavior, while the  $\tau _{21}^{-1}$ varies as $T^3 \left( e^{-E_p/k_BT} - e^{-3E_p/k_BT}\right)$ with the temperature. These new forms of temperature dependences in viscous relaxation rates have never been reported in a non-interacting $s$-wave superfluid Fermi gas where we used the same method \cite{Yavari2011}. The general trend of low temperature shear viscosity stays similar to other superfluid Fermi gases as reported in ref. \cite{Schafer2012}, and in a superfluid Fermi gas in the unitarity limit \cite{Rupak2007}, i.e. they all have divergence behavior at infinite $\beta$. This is the main result of our paper, which is important for ongoing experiential and theoretical works on ${{p}_{x}}$-wave unconventional superfluid Fermi gas in low temperature limit.
\section*{Acknowledgments}
E.D. is supported by FAPESP (Brazil) through Grant No. 2018/10813-2.


\begin{thebibliography}{00}
\bibitem {Regal2003let} C. A. Regal, C. Ticknor, J. L. Bohn, and D. S. Jin, {Phys. Rev. Lett. \textbf{90}, 053201  (2003).}
\bibitem {Regal2003nat} C. A. Regal, C. Ticknor, J. L. Bohn, and D. S. Jin, {Nature \textbf{424}, 47 (2003).}
\bibitem {Ticknor2004} C. Ticknor, C. A. Regal, D. S. Jin, and J. L. Bohn, {Phys. Rev. A \textbf{69}, 042712 (2004).}
\bibitem {Zhang2004} J. Zhang, E. G. M. van Kempen, T. Bourdel, L. Khaykovich, J. Cubizolles, F. Chevy, M. Teichmann, L. Tarruell, S. J. J. M. F. Kokkelmans, and C. Salomon, {Phys. Rev. A \textbf{70}, 030702 (2004).}
\bibitem {Schunck2005} C. H. Schunck, M. W. Zwierlein, C. A. Stan, S. M. F. Raupach, W. Ketterle, A. Simoni, E. Tiesinga, C. J. Williams, and P. S. Julienne, {Phys. Rev. A \textbf{71}, 045601 (2005).}
\bibitem {Gunter2005} K. G\"{u}nter, T. St\"{o}ferle, H. Moritz, M. K\"{o}hl, and T. Esslinger, {Phys. Rev. Lett. \textbf{95}, 230401 (2005).}
\bibitem {Gaebler2007}J. P. Gaebler, J. T. Stewart, J. L. Bohn, and D. S. Jin, {`Phys. Rev. Lett. \textbf{98}, 200403 (2007).}
\bibitem {Inada2008} Y. Inada, M. Horikoshi, S. Nakajima, M. Kuwata-Gonokami, M. Ueda, and T. Mukaiyama,
{Phys. Rev. Lett. \textbf{101}, 100401 (2008).}
\bibitem {Fuchs2008} J. Fuchs, C. Ticknor, P. Dyke, G. Veeravalli, E. Kuhnle, W. Rowlands, P. Hannaford, and C.
J. Vale , {Phys. Rev. A \textbf{77}, 053616 (2008).}
\bibitem {Nakasuji2013} T. Nakasuji, J. Yoshida, and T. Mukaiyama, {Phys. Rev. A \textbf{88}, 012710 (2013).}
\bibitem {Maier2010} R. A. W. Maier, C. Marzok, and C. Zimmermann, and Ph. W. Courteille, {Phys. Rev. A \textbf{81},
064701 (2010).}
\bibitem {Gurarie2005} V. Gurarie, L. Radzihovsky, and A. V. Andreev, {Phys. Rev. Lett. \textbf{94}, 230403 (2005).}
\bibitem {Ohashi2005} Y. Ohashi, {Phys. Rev. Lett. \textbf{94}, 050403 (2005).}
\bibitem {Ho2005} T. L. Ho and R. B. Diener, {Phys. Rev. Lett. \textbf{94}, 090402 (2005).}
\bibitem {Botelho2005} S. S. Botelho and C. A. R. S\'{a} deMelo, {J. Low Temp. Phys. \textbf{140}, 409 (2005).}
\bibitem {Iskin2005} M. Iskin and C. A. R. S\'{a} de Melo, {Phys. Rev B \textbf{72}, 224513 (2005).}
\bibitem {Iskin2006} M. Iskin and C. A. R. S\'{a} de Melo, {Phys. Rev. Lett. \textbf{96}, 040402 (2006).}
\bibitem {Cheng2005} C.-H. Cheng and S.-K. Yip, {Phys. Rev. Lett. \textbf{95}, 070404 (2005), Phys. Rev. B \textbf{73}, 064517 (2006).}
\bibitem {Levinsen2007} J. Levinsen, N. R. Cooper, and V. Gurarie, {Phys. Rev. Lett. \textbf{99}, 210402 (2007).}
\bibitem {Gurarie2007} V. Gurarie, L. Radzihovsky, {Ann. Phys. \textbf{322}, 2 (2007).}
\bibitem {Grosfeld2007} E. Grosfeld, N. R. Cooper, A. Stern, and R. Ilan, {Phys. Rev. B \textbf{76}, 104516 (2007).}
\bibitem {Iskin2008} M. Iskin and C. J. Williams, {Phys. Rev. A \textbf{77}, 041607(R) (2008).}
\bibitem {Mizushima2008} T. Mizushima, M. Ichioka, and K. Machida, {Phys. Rev. Lett. \textbf{101}, 150409 (2008).}
\bibitem {Han2009} Y.-J. Han, Y.-H. Chan, W. Yi, A. J. Daley, S. Diehl, P. Zoller, and L.-M. Duan {Phys. Rev.
Lett. \textbf{103}, 070404 (2009).}
\bibitem {Mizushima2010} T. Mizushima and K. Machida, {Phys. Rev. A \textbf{81}, 053605 (2010), \textit{ibid}., \textbf{82}, 023624 (2010).}
\bibitem {Inotani2012} D. Inotani, R. Watanabe, M. Sigrist, Y. Ohashi, {Phys. Rev. A. \textbf{85}, 053628 (2012).}
\bibitem {Inotani2015} D. Inotani and Y. Ohashi, {Phys. Rev. A \textbf{92}, 063638 (2015).}
\bibitem {Chevy2005} F. Chevy, E. G. M. van Kempen, T. Bourdel, J. Zhang, L. Khaykovich, M. Teichmann, L.
Tarruell, S. J. J. M. F. Kokkelmans, and C. Salomon, {Phys. Rev. A \textbf{71}, 062710 (2005).}
\bibitem {Jona2008} M. Jona-Lasinio, L. Pricoupenko, and Y. Castin, {Phys. Rev. A \textbf{77}, 043611 (2008).}
\bibitem {Levinsen2008} J. Levinsen, N. R. Cooper, and V. Gurarie, {Phys. Rev. A \textbf{78}, 063616 (2008).}
\bibitem {Ashurst1975} W. T. Ashurst, W. G. Hoover, {Phys. Rev. A \textbf{11}, 658 (1975).}
\bibitem {Ashurst2010} D. Baillie, P.B. Blakie, {Phys. Rev. A \textbf{82}, 033605 (2010).}
\bibitem {Enss2012} T. Enss, C. Kuppersbusch, L. Fritz, {Phys. Rev. A \textbf{86}, 013617 (2012).}
\bibitem {Schafer2012} T. Sch\"{a}fer, {Phys. Rev. A \textbf{85}, 033623 (2012).}
\bibitem {Bruun2007} G. M. Bruun, and H. Smith, {Phys. Rev. B \textbf{75}, 043612 (2007).}
\bibitem {Taylor2010} E. Taylor and M. Randeria, {Phys. Rev. A \textbf{81}, 053610 (2010).}
\bibitem {Bobrov2010} V. B. Bobrov, S. A. Trigger, and A. G. Zagorodny, Phys. Lett. A \textbf{375}, 84 (2010).
\bibitem {Enss2011} T. Enss, R. Haussmann, and W. Zwerger, {Annals Phys. \textbf{326}, 770 (2011).}
\bibitem {Bradlyn2012} B. Bradlyn, M. Goldstein, and N. Read, {Phys. Rev. B \textbf{86}, 245309 (2012).}
\bibitem {Yavari2011} H. Yavari and B. Ghafournia, {Solid State Commun. \textbf{151}, 177 (2011).}
\bibitem {Das2016} N. Das and N. Singh, Phys. Lett. A \textbf{380}, 490 (2016).
\bibitem {Darsheshdar2016ann} E. Darsheshdar, H. Yavari, and Z. Zangeneh, {Annals Phys. \textbf{370}, 105 (2016).}
\bibitem {Darsheshdar2016eur} E. Darsheshdar, H. Yavari, and S. M. Moniri, {Eur. Phys. J. Plus \textbf{131}, 178 (2016).}
\bibitem {Kagamihara2019} D. Kagamihara, D. Inotani, and Y. Ohashi, {J. Phys. Soc. Jpn. \textbf{88}, 114001 (2019).}
\bibitem {Yan2008} S. Yan and B. Yao, Phys. Lett. A \textbf{372}, 5243 (2008).
\bibitem {Mahan} G. D. Mahan, \textit{Many-Particle Physics}, {(Plenum Press, New York, 1990).}
\bibitem {Inotani2012low} D. Inotani, M. Sigrist, and Y. Ohashi, {J. Low Temp. Phys. \textbf{171}, 376(2012).}
\bibitem {Inotani2018} D. Inotani and Y. Ohashi, {Phys. Rev. A \textbf{98}, 023603 (2018).}
\bibitem{Watanabe2010} R. Watanabe, S. Tsuchiya, Y. Ohashi, {Phys. Rev. A. \textbf{82}, 043630 (2010).}
\bibitem{Schrieffer} J. R. Schrieffer, \textit{Theory of Superconductivity}, (Addison-Wesley, NY 1964).
\bibitem{Mahan1990} G.D. Mahan, \textit{Many-Particle Physics}, 2nd edition (Plenum Press, New York, 1990).
\bibitem{Rupak2007} G. Rupak, T. Schafer, {Phys. Rev. A \textbf{76}, 053607 (2007).}
\end{thebibliography}
\end{document}